\documentclass[aps,nobibnotes,twocolumn,superscriptaddress,showkeys]{revtex4-1}   	
\usepackage{graphicx}				
\usepackage{amssymb}
\usepackage{amsmath}
\usepackage{hyperref}
\usepackage{multirow}
\usepackage{placeins}

\usepackage{etoolbox}

\newtoggle{arxiv}
\togglefalse{arxiv} 

\begin{document}

\title{Boids in a Loop: Self-Propelled particles within a Flexible Boundary}
\author{A. C. Quillen}
\email{alice.quillen@rochester.edu}
\affiliation{Department of Physics and Astronomy, University of Rochester, Rochester, NY 14618, USA}
\author{J. P. Smucker}
\email{jps45@psu.edu}
\affiliation{Department of Physics and Astronomy, University of Rochester, Rochester, NY 14618, USA}
\affiliation{Department of Physics, Penn State, PA, USA}
\author{A. Peshkov}
\email{apeshkov@ur.rochester.edu }
\affiliation{Department of Physics and Astronomy, University of Rochester, Rochester, NY 14618, USA}

\begin{abstract}
We numerically explore the behavior of repelling and aligning self-propelled polar particles (boids) in 2D enclosed by a damped flexible and elastic loop-shaped boundary. We observe disordered, polar ordered (or jammed) and circulating states. The latter produce a rich variety of boundary shapes including; circles, ovals, irregulars, ruffles, or sprockets, depending upon the bending moment of the boundary and the boundary to particle mass ratio. With the exception of the circulating states with non-round boundaries, states resemble those exhibited by attracting self-propelled particles, but here the confining boundary acts in place of a cohesive force.   We attribute the formation of ruffles to instability mediated by pressure on the boundary when the speed of waves on the boundary approximately matches the self-propelled particle's swim speed.
\end{abstract}









\maketitle

\section{Introduction}
\label{sec:intro}

Active systems are non-equilibrium collections of self-propeled particles that exibit a number of striking patterns including flocking, spontaneous aggregation and formation of vortex or ring-like collective motion 
(e.g., \cite{Reynolds1987,Vicsek1995,Toner1995,Levine2000,Paxton2004,Narayan2007,Thutupalli2011}).  Inspired by biological systems exhibiting collective phenomena such as flocking \citep{Parrish1999}, artificial systems have been designed \cite{DESEIGNE:2010:ID597,PALACCI:2010:ID598,PALACCI:2013:ID684,Bricard2013} that inject energy at the microscopic level and emulate the unique properties of their biological counterparts.

Collective behaviors can also emerge in confined geometries due to 
interactions with the boundary or the surrounding fluid  (e.g., \cite{Hernandez2005,Bricard2013,Bricard2015}).
Confining walls may promote the creation of micro-scale patterns, for example wavelike cell migration modes \citep{Petrolli2019}.
Active particles can interact collectively with movable rigid or flexible objects.
For example, the fluctuations in active medium can affect the folding configurations of a flexible polymer \citep{Harder2014} while the self-propulsion energy can be harnessed to power microscopic rotating gears  \citep{Sokolov2010,Angelani2011}. 
Boundaries can be incorporated into the design of active matter based devices, for example, to generate fluid flow from confined bacteria \citep{Gao2015}. 
 For a review of active particles in crowded environments see \citet{Bechinger2016}. 
We focus here on self-propelled particles that are confined by a flexible loop-shaped boundary (e.g.,  \cite{Tian17,Nikola16,Paoluzzi2016,Deblais18,Wang19}).

Soft boundaries, including loops, membranes, thin elastic rods or plates, are interesting potential components for design. Pressure exerted by the active units can drive immersed objects to move directionally \citep{Angelani2009,Angelani2011}.  Soft boundaries can influence collective motion in active mater due to the `swim pressure' exerted by the particles on a boundary \citep{Takatori2014,Yan2015,Nikola16,Junot17}.  Flexible materials can dynamically respond with more degrees of freedom than rigid bodies such as walls, wedges or ratchets.
Such systems may have practical applications in micro bio-mechanics where flexible synthetic autonomous mechanisms  can be used as drug-delivery agents, passible cargo transport or for mechanical actuation, as suggested by \citet{Paoluzzi2016}.

In this study we numerically explore the behavior of self-propelled particles in two-dimensions that
are enclosed within a flexible circular boundary.  We search for forms of collective behavior
involving motions in the boundary, such ovals or dumbbell shapes \citep{Paoluzzi2016,Wang19} or wave-like instabilities on the boundary \citep{Nikola16}. We are interested in complex interactions between the particles and the boundary that can lead to new types of artificial mechanisms that harness collective motion.

We work with the class of Dry Aligning Dilute Active Matter which is called DADAM, (see \cite{Chate2019}).
Discrete time polar self-propelled particle models \citep{Reynolds1987,Vicsek1995}, come in deterministic or stochastic varieties (e.g., \cite{Levine2000,Chate2008,Touma2010,Henkes2011,Costanzo2018,Chate2019}) and the self-propelled particles within them are sometimes called `boids',  following \citet{Reynolds1987}.  
We focus here on the deterministic variety.    Our study is most similar to the numerical work by \citet{Nikola16,Paoluzzi2016,Wang19}
and experimental study of vibrating robotic rods by \citet{Deblais18} who also studied
 repulsive active particles in 2 dimensions that interact with a flexible boundary.  However
our simulations lack stochastic perturbations and particles within our simulations align their direction
of motion with the direction of nearby particles, as in simulations of flocking  
(e.g., \cite{Reynolds1987,Vicsek1995,Levine2000,Touma2010}).  Prior simulations of self-propelled particles
within a flexible loop have focused on non-aligning self-propelled particles with stochastically perturbed
directions of motion (e.g., \cite{Paoluzzi2016,Wang19}).

In section \ref{sec:code},
we describe our numerical model of self-propelled particles that are enclosed
inside a flexible boundary.   
In section \ref{sec:phenomena} we illustrate the phenomena seen with our simulations, 
 discuss this collective behavior and the nature of instability on the boundary.
A summary and discussion follows in section \ref{sec:sum}. Additional details for the numerical model are included in the appendix.

\begin{figure*}
\iftoggle{arxiv}{   
\includegraphics[trim=0 0 0 0,clip,width=7.0in]{fig1.png} 
}{ 
\includegraphics[trim=72 85 15 25,clip,width=7.0in]{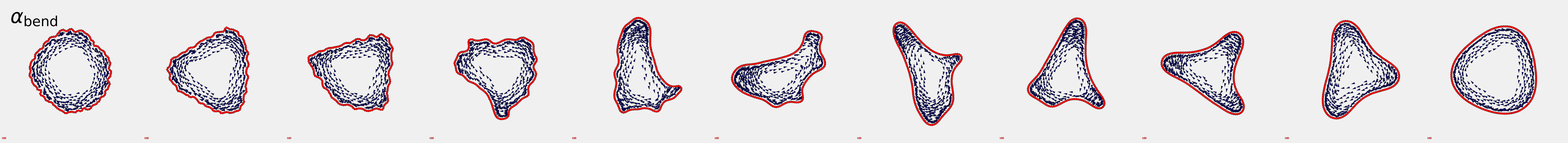} 
\includegraphics[trim=72 85 15 25,clip,width=7.0in]{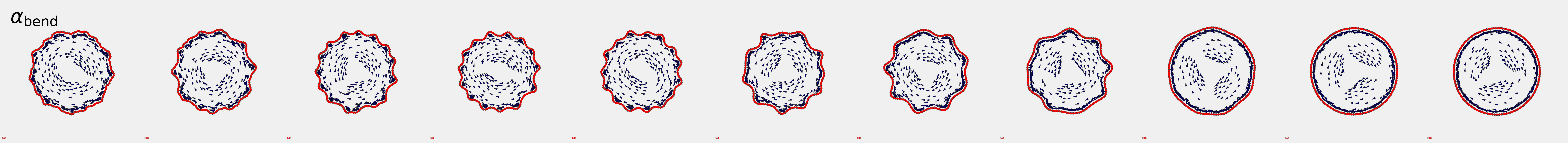} 
\includegraphics[trim=72 85 15 25,clip,width=7.0in]{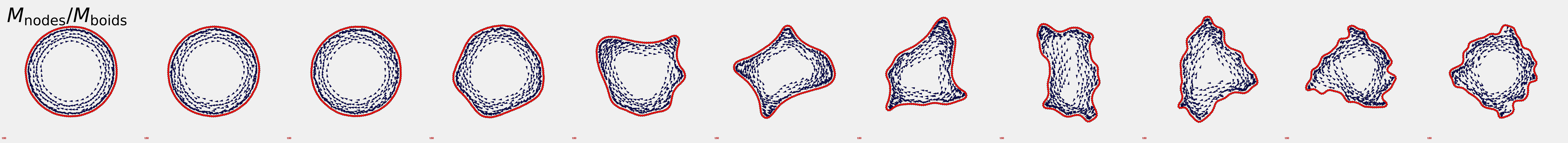} 
\includegraphics[trim=72 85 15 25,clip,width=7.0in]{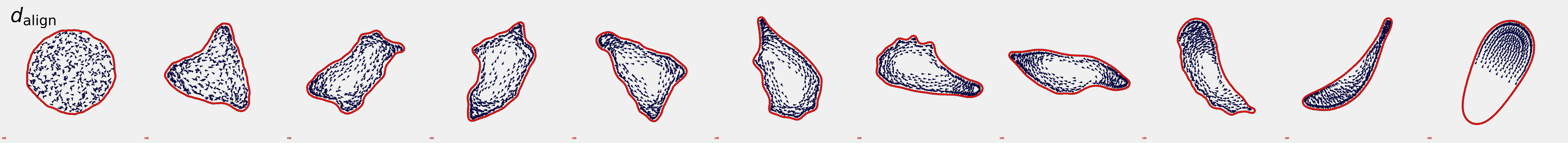} 
\includegraphics[trim=72 85 15 25,clip,width=7.0in]{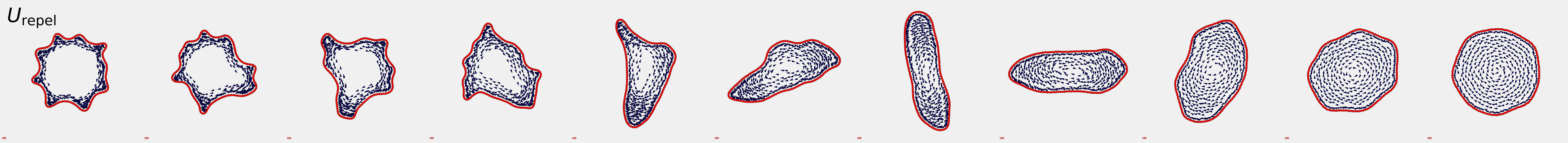} 
}
\caption{Montages of snap shots showing series of simulations with parameters listed in Table \ref{tab:common}
and \ref{tab:loops}.
In each row simulations have the same parameters except one parameter is varied.
The varied parameter is printed on the top left panel and 
increases to the right for each simulation in the row. 
a) Varying the stiffness of the boundary $\alpha_{\rm bend}$ but with a lower value of 
boid repel distance $d_{\rm repel}$.   Softer boundaries have smaller wavelength corrugations.
b) Varying bending moment, $\alpha_{\rm bend}$ but with a higher value of the repel  distance, $d_{\rm repel}$. 
c) Varying the mass of the boundary, $M_{\rm nodes}/M_{\rm boids}$.
Higher mass boundaries show smaller wavelength corrugations.
d) Varying the alignment distance $d_{\rm align}$.  A gaseous state is seen in the leftmost snapshot
and a solid-like or jammed bullet state is seen on the far right.
e) Varying the repulsion strength $U_{\rm repel}$.  
\label{fig:mont}}
\end{figure*}

\section{Boid and Boundary Model} 
\label{sec:code}

A system of self-propelled particles can be described at a fine-grained level taking into account the self-propulsion mechanism, the internal degrees of freedom of microswimmers, and the hydrodynamics. Alternatively the dynamics can be approximated via a coarse-grained approach where the motion of the self-propelled particles is described with effective forces \citep{tenHagen2015}. We adopt the second approach and neglect background hydrodynamic-like interactions. 

Our model has two particle components, a boundary that is comprised of discrete mass nodes, and a flock of self propelled particles or boids.   
We describe in detail our numerical implementation as it contains more degrees
of freedom than simulations of unconfined self-propelled particles  (e.g., \cite{Touma2010}) or self-propelled
particles with periodic boundary conditions (e.g., \cite{Gregoire2004}).

Both boundary nodes and boids can move and are massive, however boundary nodes remain in a linear chain.    Particle and node positions are denoted with ${\bf x}_i$ and the index identifies the particle or node.  
The coordinates are in two-dimensions only. The flexible boundary is initially a circular loop and encloses the boids.

We first describe the flock of boids (section \ref{sec:flock}), then the boundary (section \ref{sec:boundary}), then we discuss interactions between boids and boundary (section \ref{sec:interact}).  
Additional details on our numerical implementation are described in the appendix.
Initial conditions are described in subsection  \ref{sec:init}). The units and constraints on the time step are discussed in subsections \ref{sec:units} and \ref{sec:dt}. Additional restrictions on parameter choices are discussed
in subsection \ref{sec:other}.  The code repositories are given in subsection \ref{sec:repo}.

\subsection{The Flock of Boids}
\label{sec:flock}

A boid with index $i$ has position ${\bf x}_i^n$ at time denoted with index $n$.  The boid velocity
at the same time is ${\bf v}_i^n$ and its mass is $m_{\rm boid}$.  The total number of boids is $N_{\rm boids}$
and the total mass in boids is $M_{\rm boids} = N_{\rm boids} m_{\rm boid} $.
We update boid positions and velocities using the first order (in time) Eulerian method (as did \cite{Chate2008}) and with a fixed
time step $dt$
\begin{align}
{\bf x}_i^{n+1} &= {\bf x}_i^n + {\bf v}_i^n dt  \label{eqn:xn1} \\
{\bf v}_i^{n+1} & = {\bf v}_i^n  + \frac{dt}{m_{\rm boid}}  {\bf F}^n_i  \label{eqn:vn1} \\
 {\bf F}^n_i & = {\bf F}_{{\rm align},i}^n + {\bf F}_{{\rm repel},i}^n + {\bf F}_{{\rm interact},i}^n 
\label{eqn:Fn1}
\end{align}
where $ {\bf F}^n_i$ is a sum of forces that depend on boid position and velocity  
(${\bf x}^n_i, {\bf v}^n_i$), 
neighboring boid positions and velocities (${\bf x}^n_j, {\bf v}^n_j$ with $j \ne i$),
and nearby boundary node positions.
Hereafter we will often omit the superscript $n$.
It is useful to define a vector between two boids
${\bf r}_{ij} \equiv {\bf x}_i - {\bf x}_j$,
distance $r_{ij} = |{\bf r}_{ij}|$, and direction that is the unit vector $ \hat {\bf r}_{ij} = {\bf r}_{ij} /r_{ij}$.  

For our self propelled particles, we employ a Vicsek type of model \citep{Vicsek1995} causing nearby 
particles to align but we lack stochastic  perturbations that would change the direction of motion,  and we include
an additional inter-boid repelling force (e.g.,  as used by \cite{Levine2000,Touma2010,Henkes2011,Nikola16,Paoluzzi2016,Wang19}).  We do not apply an inter-boid attractive or cohesive force. 

The repel force on boid with index $i$ is a sum over repulsion forces
from nearby boids with index $j$
\begin{align}
{\bf F}_{{\rm repel},i} &= \sum_{i\ne j,r_{ij}<2d_{\rm repel} } \frac{m_{\rm boid} U_{\rm repel}}{d_{\rm repel}} e^{-r_{ij}/d_{\rm repel}}  \hat {\bf r}_{ij}.
\end{align}
Here $U_{\rm repel}$ has units of the square of velocity and $d_{\rm repel}$ characterizes the scale of the repulsive interaction.
We only apply the repel force for boid pairs separated by $r_{ij}<2d_{\rm repel}$.
The repel force is applied equally and oppositely to boid pairs.
This repel force is exponential (as was that adopted by \cite{Touma2010}).  We also explored a repel force proportional to the inverse interboid distance and saw similar collective phenomena.

An align or steer force also serves to propel the boids at a velocity that is approximately $v_0$.
The align or steer and self-propelling force exerted on boid $i$ is
\begin{align}
{\bf F}_{{\rm align},i} &=   \alpha_{\rm align} m_{\rm boid} (v_0 \hat {\bf w}_i  - {\bf  v}_i)  \label{eqn:align}  \\
\hat {\bf w}_i & =  \frac{{\bf w}_i}{|{\bf w}_i|}.
\end{align}
Here $\alpha_{\rm align}$ has units of inverse time and $v_0$ is the boid speed,
equal to the `terminal velocity' in the model by \citet{Touma2010}.
The unit vector $\hat {\bf w}_i$
 is multiplied by $v_0$ so that the boid accelerates if its speed is slower than $v_0$
 and it decelerates if it is going faster than $v_0$.  
 A distance $d_{\rm align}$  characterizes the scale of the alignment interactions.
A boid lacking neighbors that are within alignment distance $d_{\rm align}$ is propelled 
using the boid's own current velocity direction with $ {\bf w}_i =  {\bf v}_i$.  
For a boid with near neighbors,
the vector $ {\bf w}_i$ is computed from the velocities of nearby boids, similar
to prior numerical models  \citep{Vicsek1995,Levine2000},
\begin{equation}
{\bf w}_i =   \sum_{i\ne j, r_{ij} < d_{\rm align}}  {\bf v}_j  .
\end{equation}

%

\subsection{The Flexible Elastic Boundary}
\label{sec:boundary}

The numerical description of our flexible boundary is similar to that used by  \citet{Nikola16}
(see VI  of their supplements). 
The boundary is described with a chain of mass nodes, each of mass $m_{\rm node}$. Each node is initially separated from its two nearest neighbors by a distance $\Delta s$. The chain is closed by connecting its two endpoints so that it forms a loop.
A node at position ${\bf x}_i$ has neighbors ${\bf x}_{i+1}$ and ${\bf x}_{i-1}$ with indices given modulo the total number of nodes in the chain, $N_{\rm nodes}$.
The total mass in the boundary is $M_{\rm nodes} = N_{\rm nodes} m_{\rm node}$.
To maintain boundary length, each consecutive pair is separated by a spring with rest length $\Delta s  = 2 \pi R/ N_{\rm nodes}$ where $R$ is the initial loop radius.    Using a thin elastic beam approximation, we apply forces to the nodes that allow the boundary to resist bending.  We first discuss the bending forces and then the spring forces.

We update node positions and velocities using equations \ref{eqn:xn1} and \ref{eqn:vn1} but with
$m_{\rm boid}$ replaced with $m_{\rm node}$. 
Instead of equation \ref{eqn:Fn1}, the sum of forces on node $i$ at time step $n$ is  
\begin{align}
 {\bf F}^n_i & = {\bf F}_{\rm bend,i}^n + {\bf F}_{\rm spring,i}^n + {\bf F}_{{\rm interact},i}^n 
  + {\bf F}_{\rm damp,i}^n
\end{align}
and the forces depend on node positions and velocity  
(${\bf x}^n_i, {\bf v}^n_i$), 
neighboring node positions and velocities (${\bf x}^n_j, {\bf v}^n_j$ with $j \ne i$),
and nearby boid positions.


The Euler-Bernoulli theory of thin elastic beams describes the centerline
of a beam with a curve ${\bf X}(s)$ where $ds$ gives length along the boundary.
The elastic potential energy depends on 
\begin{equation}
 U_{\rm bend} = \int ds  \frac{\alpha_{\rm bend}}{2} \left( {\bf X''}(s) \right)^2 \end{equation}
where ${\bf X}'' = \frac{\partial^2 {\bf X}(s)}{ \partial s^2} $ is the curvature.
The coefficient  $\alpha_{\rm bend} = EI$, is known as the bending moment or flexural rigidity,  with $E$
 the elastic modulus and $I$ is the second moment of area integrated on the beam's cross-section. 
For a linear beam  oriented on the $x$ axis with linear mass density $\mu$, and displacement from the
$x$ axis $w(x,t)$,
 the above potential energy gives equation 
of motion 
\begin{equation}
 \mu \frac{\partial^2 w}{\partial t^2} =  -\alpha_{\rm bend} \frac{\partial^4 w}{\partial x^4} . 
\end{equation}

We discretize our boundary by putting its mass into a consecutive set of mass nodes ${\bf x}_i$, each separated by distance $\Delta s$. The curvature at a node  
\begin{equation}
{\bf x}_i'' \approx (\Delta s)^{-2} (  {\bf x}_{i+1}  +  {\bf x}_{i-1}  -2{\bf x}_{i}). 
\end{equation}
The potential energy for the discrete chain  
\begin{equation}
 U_{\rm bend} = \sum_i\    \frac{\alpha_{\rm bend} }{(\Delta s)^3} (3 |{\bf x}_i|^2 +  {\bf x}_i \cdot  {\bf x}_{i+2} - 4 {\bf x}_i \cdot {\bf x}_{i+1}  ) .
 \end{equation}
Taking the derivative of potential energy $U$ with respect to node position ${\bf x}_i$ gives the force on a node
\begin{align}
  {\bf F}_{{\rm bend},i}  &= -\frac{\partial U}{\partial {\bf x}_i} \nonumber \\
  & = -\frac{\alpha_{\rm bend} }{(\Delta s)^3} ({\bf x}_{i-2}-4 {\bf x}_{i-1} + 6{\bf x}_{i} - 4 {\bf x}_{i+1} + {\bf x}_{i+2} ) .
\end{align}
The equation of motion 
\begin{align}
m_{\rm node} \frac{d^2{\bf x}_i}{dt^2} =  - \frac{\alpha_{\rm bend}}{(\Delta s)^3} ({\bf x}_{i-2}-4 {\bf x}_{i-1} + 6{\bf x}_{i} - 4 {\bf x}_{i+1} + {\bf x}_{i+2} ) \label{eqn:BB}
\end{align}
 is a discrete approximation to the equation of motion from Euler-Bernoulli elastic beam theory (e.g., \cite{snakes,bergou08}).

We insert a spring between each consecutive node on the boundary.
The springs are intended to  maintain a nearly constant length boundary.
The total potential energy due to springs is
\begin{equation}
U_{\rm spring} = \sum_i \frac{k_s}{2} ( r_{i,i-1}   - \Delta s )^2
\end{equation}
where  $r_{i,i-1} = | {\bf x}_i - {\bf x}_{i-1}|$ is the distance between two consecutive nodes,
$\Delta s$ is the rest spring length and $k_s$ the spring constant.
The force exerted on each node due to the springs is 
\begin{align} {\bf F}_{{\rm spring},i} &= 
-  k_s \frac{({\bf x}_i - {\bf x}_{i-1} ) }{ r_{i,i-1}}  (  r_{i,i-1} - \Delta s)  \nonumber \\
& \qquad
-  k_s \frac{({\bf x}_i - {\bf x}_{i+1} ) }{ r_{i,i+1}}  (  r_{i,i+1} - \Delta s) .
\end{align}
This  follows common implementations of N-body mass/spring models (e.g., \cite{Frouard2016}).


To mimic an external viscous or friction like boundary interaction, we add a velocity dependent
damping force on each boundary node
\begin{equation}
 {\bf F}_{{\rm damp},i} = - m_{\rm node} \gamma_{\rm damp} {\bf v}_i, 
 \end{equation}
where damping parameter $\gamma_{\rm damp}$ is in units of inverse time
and ${\bf v}_i$ is velocity of the node.

\subsection{Boundary Node/Boid interactions}
\label{sec:interact}

We apply an equal and opposite repulsive force to each pair of boundary and boid particles.
The force on particle $i$ (either a boundary node or boid) from particle with index $j$ (of the opposite type)
\begin{equation}
{\bf F}_{{\rm interact},i}   =  \sum_{j, r_{ij} < 3 d_{\rm interact}}  F_{\rm interact} e^{-r_{ij}/d_{\rm interact}} \hat {\bf r}_{ij}.
\end{equation}
The distance $d_{\rm interact}$ describes the range  of the interaction.
We only apply the force  at distances  $r_{ij} < 3 d_{\rm interact}$.
The parameter $F_{\rm interact}$ determines the strength of the interaction.
As long as the interaction force
causes accelerations that exceed those from other forces and so 
causes reflection off the boundary faster than interboid distance
travel times, the collective behavior should not be sensitive to 
$d_{\rm interact}$ or $F_{\rm interact}$.

\section{Collective phenomena} 
\label{sec:phenomena}

In Figure \ref{fig:mont}, each row shows a series of 11 simulations. 
Each panel is a simulation snap shot that shows the boid distribution and boundary morphology at
the end of a simulation. Boundary particles are shown in red.
Each boid is marked with a navy blue isosceles triangle.  The vertex with narrowest angle marks the
direction of motion. In each simulation series, parameters are identical
except for one parameter which is consecutively increased in each simulation. 
Common parameters for these simulations are listed in Table \ref{tab:common}.
Additional parameters  for the series of simulations are listed in Table \ref{tab:loops}. 
These series have been done with $N_{\rm boids} = 400$, however we saw similar 
phenomena with $N_{\rm boids} = 100,200$ and 800.
A live animation showing a circulating state can be seen here \url{https://aquillen.github.io/boids_in_a_loop/}.
This animation is part of the first series of simulations and has bending moment 
$\alpha_{\rm bend}/(M_{\rm boids} v_0^2 R) = 10^{-3}$.
The 5-th panel (from the left) in Fig \ref{fig:mont}a, the 7-th panels in Fig \ref{fig:mont}c and d and
the 4-th panel in Fig \ref{fig:mont}e all have parameters approximately the same
as this animation.

Below we describe the different types of boid and boundary behavior seen in our simulations.
In section \ref{sec:phase}, we discuss divisions in parameter space that separate  gaseous, circulating and jammed states.
In section \ref{sec:sens},
we discuss the sensitivity of  boundary morphology to  simulation parameters.
In section \ref{sec:inst}, we discuss the nature of the instability that causes the boundary to be
ruffled or corrugated.

\begin{figure}
\includegraphics[trim=0 10 5 15,clip,width=3.0in]{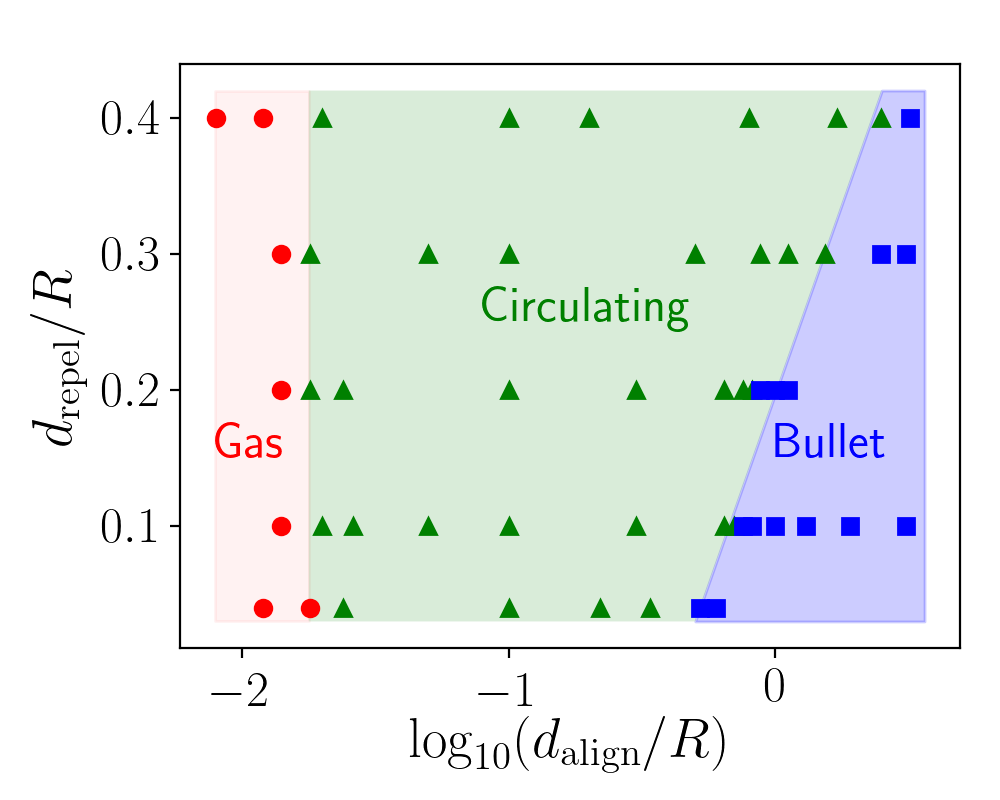} 
\includegraphics[trim=0 10 5 15,clip,width=3.0in]{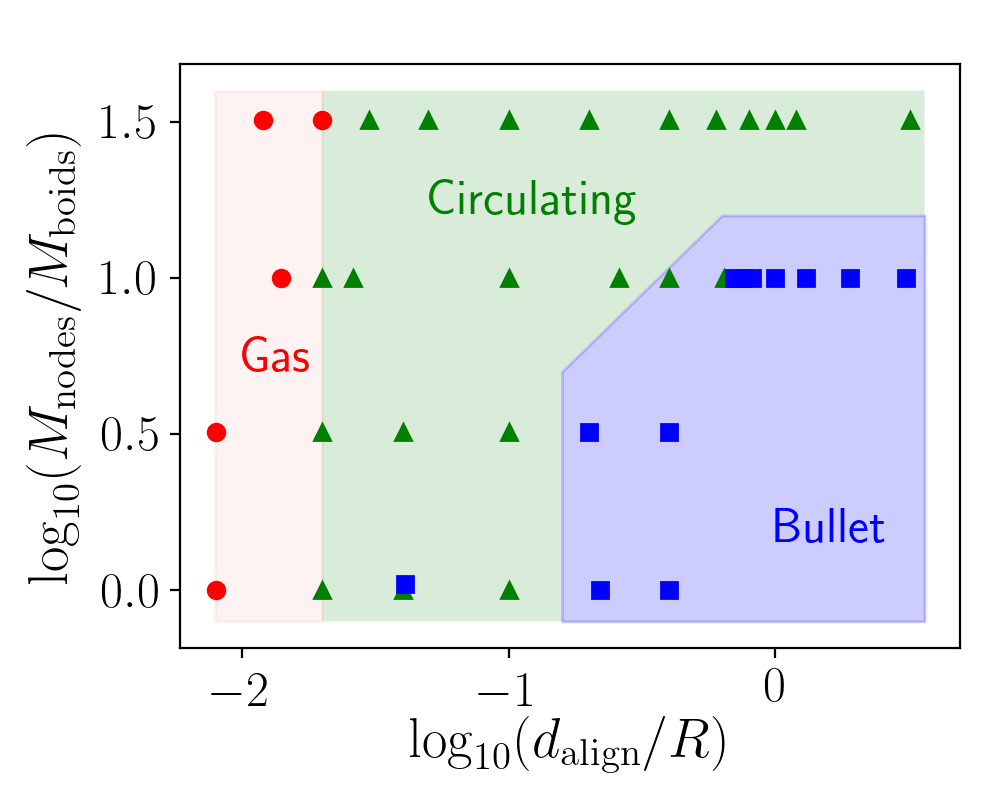} 
\includegraphics[trim=0 10 5 15,clip,width=3.0in]{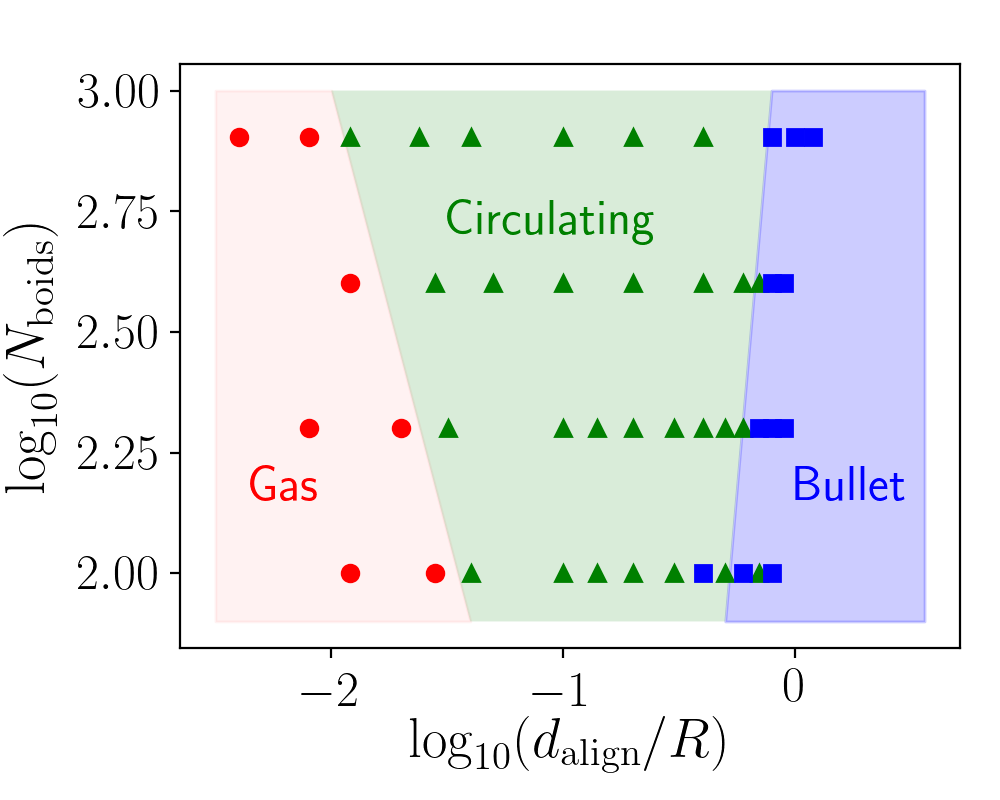} 
\caption{Dependence of the type of collective motion on  align distance and other parameters.
a) Phases are plotted as a function of the repel  and align distances.
b) Phases are plotted as a function of the ratio of boundary and boid mass and the align distance.
c) Phases are plotted as a function of the number of boids  and the align distance.
Red circles represent simulations giving gaseous states, green triangles represent those
giving circulating states, and blue squares are those that ended in jammed or bullet states.
We have roughly shaded the different regions.  The jammed bullet phase is present when
the alignment is strong, whereas the gaseous state is present when the alignment is weak.
Circulating states lie in between the gaseous and bullet states.  
The simulations used to make this figure have parameters listed in Table \ref{tab:common} and the rightmost columns of 
Table \ref{tab:loops}.
\label{fig:threephase}}
\end{figure}

We see three types of collective phenomena, a disordered gaseous state, a solid-like state and 
rotating or circulating states.

We first discuss the disordered {\bf gaseous} state.  Boids are not aligned with each other, there is little circulation or rotation and the boid velocity dispersion is high.
This state is characterized by a weak or short range alignment force.   An example of this state is
in the leftmost panel of  Figure \ref{fig:mont}d (fourth row from top).  This particular simulation has  a very short 
alignment distance, $d_{\rm align} = 0.01 R$.  Numerically we find that $d_{\rm align} \alpha_{\rm align}/v_0  \lesssim 0.01$ gives a gaseous state.
We see disordered gas-like behavior with little to no align forces, as encountered in simulations of 2-dimensional swarms of unconfined unipolar self-propelled particles, \citep{Levine2000,Gregoire2004,Touma2010,Chate2019}.  
Our model lacks stochastic perturbations. However, billiards within in a non-round but convex boundary can be chaotic \citep{Bunimovich1979}.  Even if our boundary was smooth instead of comprised of discrete nodes, ergodic behavior can be introduced via boids reflecting off the boundary.  Ergodic behavior would also be introduced by the interboid repulsion forces as  interactions occur frequently because the boids are confined.   

We also see a  solid-like jammed {\bf bullet} state.   Here all boids are moving in the same direction.  
Boid positions and velocities appear frozen in a frame moving with along with them.   
The boid velocity dispersion is low and boids do not move relative to each other.  
This state is characterized by a strong or long range alignment force and a lower mass
boundary that is easily pushed by the boids.   
A low damping rate on the boundary aids in forming this state.
An example of this state is in the rightmost panel of  Figure \ref{fig:mont}d (fourth panel from top)
with $d_{\rm align}/R = 1.1$.  Numerically we find that this state is likely  when  
 $d_{\rm align} \alpha_{\rm align}/v_0  \gtrsim 1$.
Even though our simulations lack an interboid attractive force,  confinement caused by the boundary can cause a jammed state.  
This state is similar to the jammed state seen previously in simulations of confined soft repelling self-propelled particles at high density \citep{Henkes2011}. 
Like ours, the simulations by \citet{Henkes2011} lack an alignment force, however  their  
boundary was rigid.
The jammed state is perhaps also similar to moving cohesive groups or droplet states seen in simulations of unconfined unipolar self-propelled particles that attract each other (e.g., \cite{Gregoire2004,Touma2010}).

Lastly we also see
  rotating or {\bf circulating} states. 
The boids are circulating within the boundary.    
The boundary can be rotating but is usually moving more slowly 
than the boids which all circulate in the same direction.   The boundary shape can be circular, oval, irregular or
sprocket shaped.   Oval loop-shaped flexible boundaries were previously
 seen in simulations of non-aligning self-propelled particles  \citep{Paoluzzi2016,Wang19}. 
We use the word `sprocket' rather than `gear' or `ratchet' to describe states with more than a few
radial projections. A sprocket is usually used to engage a chain and is distinguished from a gear in that sprockets are never meshed together.  A `ratchet' is part of a mechanical device used for turning objects that allows continuous linear or rotary motion in only one direction.  

For the irregular and sprocket shapes, the boundary is  deformed by groups of boids.  As the boids circulate, bulges in the boundary travel along the boundary.  Irregular or sprocket boundaries are more likely if
the boundary mass exceeds the total boid mass but the boundary is not so massive
that the boids cannot push it.  Irregular or sprocket boundaries are more likely with a more flexible rather than stiff boundary.  As is true for the bullet states, the circulating states arise in the absence of interboid attraction. 
The confining boundary serves in place  of attractive forces that cause circulating states in unconfined self-propelled particles (e.g., \cite{Touma2010}).  Because there is no attraction force between boids, we do not see multiple separate flocks, though we do see
clumps of boids in divots or pockets moving along the boundary.


Long-lived states can depend on the initial boid velocity distribution.  When alignment is strong and the boundary is lower mass, initially rotating boids are less likely 
to go into the bullet state.  Once a system goes into a bullet state, we find that it
stays there.   Circulating states can nevertheless be long lived and even after long integrations, 
with $t_{\rm max} >100 t_R$, the simulation won't fall into a bullet
state even if a different initial velocity distribution would put the system in such a state.


The most interesting of the states seen in our simulations are those where the  boundary becomes corrugated. \citet{Sokolov2010} and \citet{DiLeonardo2010} describe an asymmetric rigid nano-fabricated gear that is spun by bacteria.  In contrast, here we find that a flexible loop-shaped boundary can become corrugated and the corrugations can rotate because of unipolar self-propelled particles that are moving within the boundary.  
We could be seeing a modulational instability due to swim pressure inhomogeneities near the boundary that was predicted for non-aligning self-propelled particles by \citet{Nikola16}.

Increased boid density near the boundary (bordertaxis) is particularly noticeable in the simulation with
higher repel distance $d_{\rm repel}$, (Figure \ref{fig:mont}b or second row).
The interplay of self-propulsion, confinement and stochastic processes is often sufficient to
explain accumulation of self-propelled particles on or near a boundary \citep{Elgeti2013,Fily2014,Ezhilan2015,Paoluzzi2016,Caprini2018,Deblais18,Wang19}.
Here we lack stochastic perturbations, however boundary-boid and boid-boid interactions  
serve as a source of chaotic behavior that might aid in increasing the boid density near the boundary
via diffusive-like behavior.  
Boids on the boundary only feel repulsion from other boids 
on one side allowing them to be closer together than boids in the interior.  




\subsection{Phase diagrams}
\label{sec:phase}

In Figure \ref{fig:threephase},
we show phase plots delineating gaseous, circulating and bullet states.
Figure \ref{fig:threephase}a shows phases as a function 
of repel and alignment distances,  $d_{\rm repel}$ and $d_{\rm align}$. 
Figure \ref{fig:threephase}b shows phases 
as a function of total boundary to boid mass ratio and the  align distance
and Figure \ref{fig:threephase}c shows phases as a function of the number of boids and the align distance. 
For this last figure we set $d_{\rm repel} \propto \sqrt{N_{\rm boids}}$ so that the repel distance divided
by mean boid number density remains constant in the different simulations.  
Otherwise the high number density simulations would
be at high pressure as boid repulsion would be pushing them up against the boundary.   

The  parameters for the
 simulations shown in  Figure \ref{fig:threephase} 
 are listed in Table \ref{tab:common} and in the rightmost  columns  in Table \ref{tab:loops}.
In Figure \ref{fig:threephase} red circles represent simulations giving gaseous states, green triangles represent those
giving circulating states, and blue squares are simulations that ended in bullet states.
Classification for this plot was done by eye from simulations run in the browser.
We have shaded the different regions to show the locations of the different phases.

The transition between circulating and gaseous  states is primarily sensitive to the align force strength and distance and the boid number density.
The gas/circulating phases dividing line on Figure \ref{fig:threephase}c has slope consistent with alignment distance proportional to the mean distance between boids or $ d_{\rm align}  \propto 1/\sqrt{N_{\rm boids}}$.
If the boid number density is higher, a smaller alignment distance allows them to circulate. 

The transition line between bullet and circulating states is sensitive to a number of parameters.
More flexible, less damped and lower mass boundaries are more likely to  elongate and trap boids, aiding in formation of a jammed state.  
Confined self-propelled soft particles at high density jam \citep{Henkes2011}, and unconfined self-propelled particle with strong cohesion can form moving 
solid-like droplets \citep{Gregoire2004,Touma2010}.  
The sensitivity of the bullet/circulating phase line to bending moment $\alpha_{\rm bend}$, damping parameter $ \gamma_{damp}$ and  mass ratio   $M_{\rm nodes}/M_{\rm boid}$    would be consistent with a picture where strong alignment pushes the boids into the boundary, increasing their density, but where the jammed state is only maintained when the boundary can fold and trap them.

\subsection{Sensitivity of boundary corrugations on simulation parameters}
\label{sec:sens}

We discuss the 5 series of simulations shown in Figure \ref{fig:mont} and with parameters 
listed in Tables \ref{tab:common} and \ref{tab:loops}.  
In  Figure \ref{fig:mont}a (top panel) we show a series of simulations, all with the same
parameters except that bending moment $\alpha_{\rm bend}$ increases from simulation to simulation.
The factors used to increase the varied parameter, here $\alpha_{\rm bend}$,
 in each series are also listed in Table \ref{tab:loops}.
The varied parameter is computed as follows.
The lowest  value of  $\alpha_{\rm bend}/(M_{\rm boids} v_0^2 R)$ in the first series is $10^{-4}$.  The factor used
to vary this parameter is 1.7. The 11-th simulation has bending  moment 
$\alpha_{\rm bend}/(M_{\rm boids} v_0^2 R) = 10^{-4} \times  (1.7)^{10} = 0.02$.
This set of simulations has $d_{\rm repel}=0.1$ so has a fairly short range repulsive force.
With a very flexible boundary (on the left in Figure \ref{fig:mont}a) 
and small $\alpha_{\rm bend}$, the boundary has many  corrugations.
As the bending moment increases, the wavelength of the boundary corrugations increases. 

The second series of simulations  shown in Figure \ref{fig:mont}b  (second row)
is similar to the first series except the repel distance $d_{\rm repel}=0.35$ is larger.
The repel distance is large enough that boids are pushed against the boundary by their repulsion alone.
This differs from the simulations 
at lower $d_{\rm repel}$ where only the centrifugal force due to  their circulation pushes
them up against the boundary.
Despite being in  a different regime, we also see boundary corrugations in the series shown in Figure \ref{fig:mont}b,
and again with  wavelength increasing with increasing bending moment.
In this regime a single angular Fourier mode  often dominates, whereas
at lower repel distance $d_{\rm repel}$  the boundary corrugations were more irregular.
With higher $d_{\rm repel}$ and lower bending moment $\alpha_{\rm bend}$, the boundary looks like a sprocket or a gear.

We were most surprised by 
the third series of simulations, shown in Figure \ref{fig:mont}c  (third row).
In this series of simulations, the boundary mass is increased, with low mass boundaries on the left
and high mass  boundaries on the right.
We had expected that a lower mass boundary would show more corrugations because it would be easier
for the boids  to push the boundary.  However,  we find that the opposite is true; 
the higher mass boundaries have boundaries with more corrugations.

In Figure \ref{fig:mont}d (fourth row), we vary the alignment distance $d_{\rm align}$.
This set of simulations shows the transition from a gas-like state, at low $d_{\rm align}$ on the left 
to the jammed bullet-like state at high $d_{\rm align}$, on the right.
In some of the intermediate simulations we saw a circulating flock of boids that moved back and forth from one side of a boundary to the other.

In Figure \ref{fig:mont}e (fifth row), we vary the repel force strength $U_{\rm repel}$.
This parameter affects the boid density.  We find that the boundary is more likely to be corrugated
when the  boid density  is higher near the boundary and at  lower repel strength, $U_{\rm repel}$.

\begin{table}
\caption{Common parameters for simulation series \label{tab:common}}
\begin{tabular}{llllllll}
\hline
 $N_{\rm nodes}$                   & 150    \\
 $\alpha_{\rm align} t_R$       & 3   \\
 $\gamma_{\rm damp} t_R$    &0.1  \\
 $k_s  m_{\rm node}^{-1} t_R^2$ &  $2 \times 10^{4} $  \\
 $F_{\rm interact}  M_{\rm boids}^{-1} v_0^{-2} R$ & 1.5 \\
 $d_{\rm interact}/R$               & 0.02  \\
 $dt/t_R$                                 & 0.005   \\
 $t_{\rm max}/t_R$                 &  50  \\
 $\epsilon_{ks}$ & 0.03  \\  
 \hline
\end{tabular}
\\
{The parameter $\epsilon_{ks}$ is defined in equation \ref{eqn:epsilon_ks}.
}
\end{table}

\begin{table*}
\caption{Simulation series \label{tab:loops}}
\begin{tabular}{p{2.8cm} p{1.6cm} p{1.6cm} p{1.6cm} p{1.7cm} p{1.6cm} p{1.7cm} p{2.2cm} p{2.2cm} }
\hline
Varying  & \mbox{bending} \mbox{moment} & \mbox{bending} \mbox{moment} & 
\mbox{boundary} \mbox{mass} & \mbox{align} \mbox{distance}  & \mbox{repel} \mbox{strength}
& \mbox{align+repel} \mbox{distances} & \mbox{align distance}, \mbox{boundary mass}  & \mbox{align distance},
 \mbox{boid number}\\
\hline 
Figure                        & \ref{fig:mont}a &\ref{fig:mont}b&\ref{fig:mont}c & \ref{fig:mont}d &  \ref{fig:mont}e & \ref{fig:threephase}a & \ref{fig:threephase}b & \ref{fig:threephase}c \\
Factor                         & 1.7   & 1.7   & 1.5 & 1.6 & 1.5 & - & - & - \\
$\alpha_{\rm bend}/(M_{\rm boids} v_0^2 R)$ 
                                       &$[10^{-4} , 0.01]$ &$[10^{-4} , 0.01]$ & $10^{-3}$ & $10^{-3}$ & $10^{-3}$ &  $10^{-3}$ &  $10^{-3}$ & $10^{-3}$ \\
 $M_{\rm nodes}/M_{\rm boids}$  & 10 &10 & [1,57] &10 & 10  & 10 & [1,32] & 10\\
$d_{\rm align}/R$              & 0.2& 0.2 &0.2 & [0.01,1.1]  & 0.2 & [0.01,3.3] & [0.01,3.3] & [0.01,1.3] \\
  $U_{\rm repel}/v_0^2 $   &   0.1  & 0.1 & 0.1& 0.1& [0.03,1.6] & 0.1 &0.1 &0.1  \\
   $d_{\rm repel}/R$                 &0.1 & 0.35 & 0.1 & 0.1 & 0.1 & [0.04,0.4]& 0.1 
   								& 0.1  $\sqrt{\frac{N_{\rm boids}}{400}} $\\
   $N_{\rm boids}$                    & 400   & 400 & 400 & 400 & 400 &  400 & 400 & [100,800] \\
 Initial conditions          & rotating & rotating & rotating & not rotating & rotating  & not rotating  & not rotating & not rotating \\
 \hline
\end{tabular}
\\
{ 
The first  row gives the parameter or parameters varied for the series.
Each column gives parameters for simulations that are shown in the Figure listed
in the second row of the table.  Additional parameters for these simulations are listed in Table \ref{tab:common}.
Numbers in brackets give the range for the parameter that is varied. 
The third row, labelled `Factor' gives the multiplicative factor used to increase the varied parameter
for each consecutive simulation in Figure \ref{fig:mont}.
}
\end{table*}

\subsection{Instability on the boundary}
\label{sec:inst}

Prior studies have described the types of collective motion as phases
and delineated boundaries between these phases in parameter space, similar to phase transitions (e.g., \cite{Vicsek1995,Touma2010,Henkes2011}).  The higher number of free parameters present in our
system and sensitivity to initial conditions makes it more challenging to delineate transitions between gas-like, solid-like and circulating collective motion. 
The most novel phenomena illustrated by our dynamical system is corrugations
in the boundary that grew during the simulations. The dynamics of the boundary is coupled to the collective motions.
  Instead of examining in more detail the sensitivity of
the gas-like/circulation and circulation/bullet phases to system parameters, we examine
the nature of the instability leading to the growth of corrugations on the boundary.  

Hydrodynamic analogies for our boundary corrugations include ripples excited
on a flag by wind, or the Kelvin-Helmholtz instability which is driven by the
velocity difference across an interface between two fluids.
Classically, instabilities can be studied by linearizing equations of motion
and deriving a dispersion relation for wave-like solutions.  The dispersion relation relates a the frequency of oscillation to a wavevector.  
Frequencies that have complex parts  when the wavevectors are real, correspond to wavelengths that are unstable to amplitude growth.

Using Euler-Bernoulli theory,  the wave equation for a linear elastic beam under tension and with an applied force
\begin{equation}
\mu \frac{\partial^2 w}{\partial t^2} =  -\alpha_{\rm bend} \frac{\partial^4 w}{\partial x^4}  + T \frac{\partial^2 w}{\partial x^2}
+  p(x)  \label{eqn:waveeq}
\end{equation}
where $w(x)$ is beam displacement,  $\alpha_{\rm bend}$ is the bending moment
or flexural rigidity, $\mu$ is the beam's linear mass density, and $p(x)$ is an applied force per unit length.
We can use this equation to model the dynamics of our flexible boundary.  
Here the horizontal coordinate $x$ is a plane parallel
approximation to 
 $R \theta$ in polar coordinates along the boundary with periodic boundary conditions and $w(x)$ is a radial
displacement of the boundary away from its rest, circular state.
As discussed previously,
the linear mass density in the boundary $\mu = M_{\rm nodes}/(2\pi R)$.
In equation  \ref{eqn:waveeq}
we have included a term dependent upon tension $T$, the longitudinal tension in the boundary.
We estimate a mean value for the tension using equation \ref{eqn:tension} and depending upon
the total boid mass and associated pressure. 
The applied force $p(x)$ we assume is due to boids pushing up against the boundary.
We refer to this applied force as `swim pressure' (following \cite{Takatori2014,Yan2015,Nikola16}) or `boid pressure'.  

A perturbative solution of equation \ref{eqn:waveeq} with displacement
$w(x,t) \propto e^{i(\omega t - kx)}$,  frequency $\omega$ and wavevector $k = 2 \pi/\lambda$,  for  wavelength $\lambda$, and with applied pressure $p=0$, gives  a dispersion relation 
\begin{equation}
 \omega^2 = \frac{\alpha_{\rm bend}}{\mu} k^4 +  \frac{T}{\mu} k^2 . \label{eqn:disp}
\end{equation}
The tension related $k^2$ and bending rigidity related $k^4$ terms are consistent with discussion
on active particle mediated boundary instability by \citet{Nikola16}. 

If the boids are moving parallel to a straight surface, they will not interact with the boundary.
However if they are moving next to a curved surface their trajectories must curve.
The pressure on the boundary due to the boids depends on the curvature of the boundary
and the boid density 
$p_{\rm swim} \propto  \rho_{\rm boid} \frac{\partial^2 w}{\partial x^2}$.    The pressure force is opposite
 that due to tension in the boundary, as it would push in the same 
direction as a bulge in the boundary, rather than counter it. In this sense, the boid swim pressure
acts like pressure variations in an incompressible fluid near a boundary 
that is derived from linearization of Bernoulli's equation.
We estimate the pressure on the boundary 
\begin{equation}
p_{\rm swim} \sim -\beta_{\rm swim}  M_{\rm boid} \frac{v_0^2}{2 \pi R} \frac{\partial^2 w}{\partial x^2} . \label{eqn:pswim}
\end{equation}
where $\beta_{\rm swim} $ is a dimensionless factor that we can adjust.  This gives a simple approximate model
for variations in boid pressure exerted along a corrugated boundary and is in a similar form to
that predicted in equation 27 by \citet{Nikola16}.
This form for the swim pressure gives a term in the wave equation similar to the tension term (see equation \ref{eqn:tension} for tension) but with the opposite sign (and this is also consistent with the discussion by \citet{Nikola16} in their supplements). The dispersion relation (in equation \ref{eqn:disp}) becomes 
\begin{equation}
 \omega^2 = \frac{\alpha_{\rm bend}}{\mu} k^4 +  \frac{T(1 - \beta_{\rm swim})}{\mu} k^2 . \label{eqn:disp2}
\end{equation}

\begin{figure}
\includegraphics[trim=25 5 0 5,clip,width=3.0in]{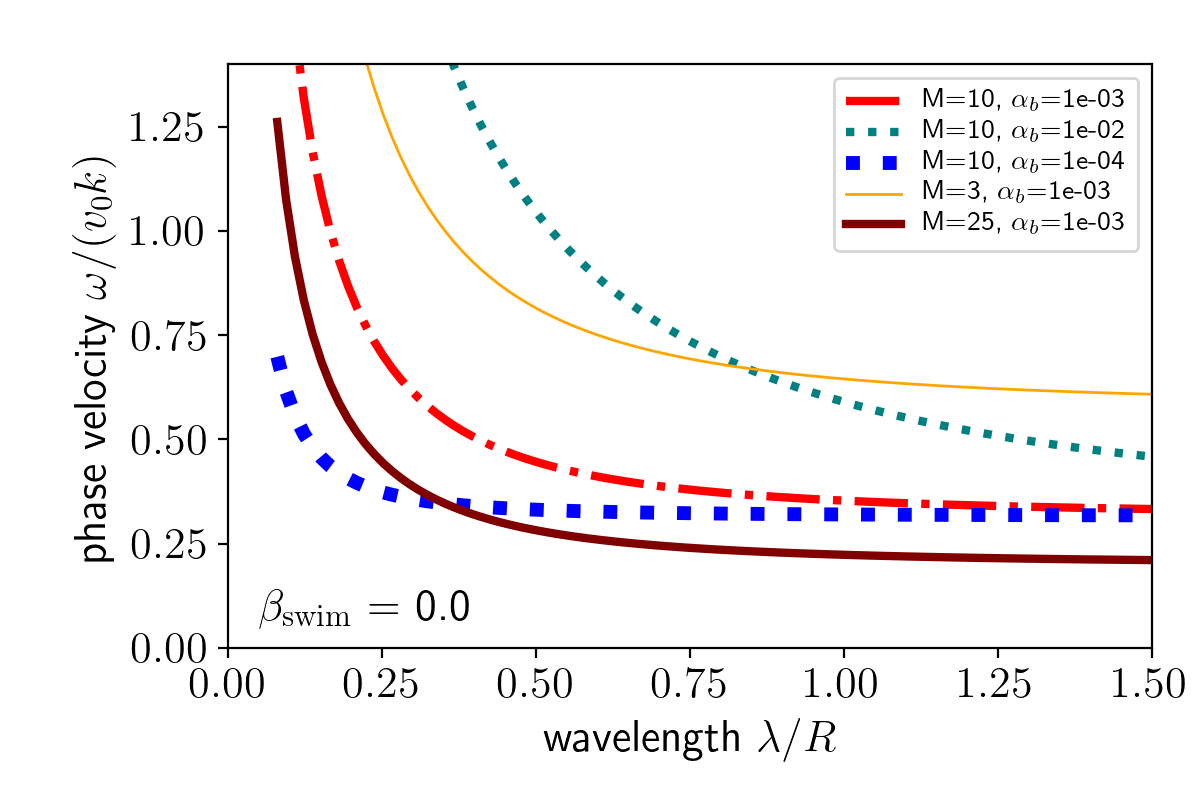} 
\includegraphics[trim=25 5 0 5,clip,width=3.0in]{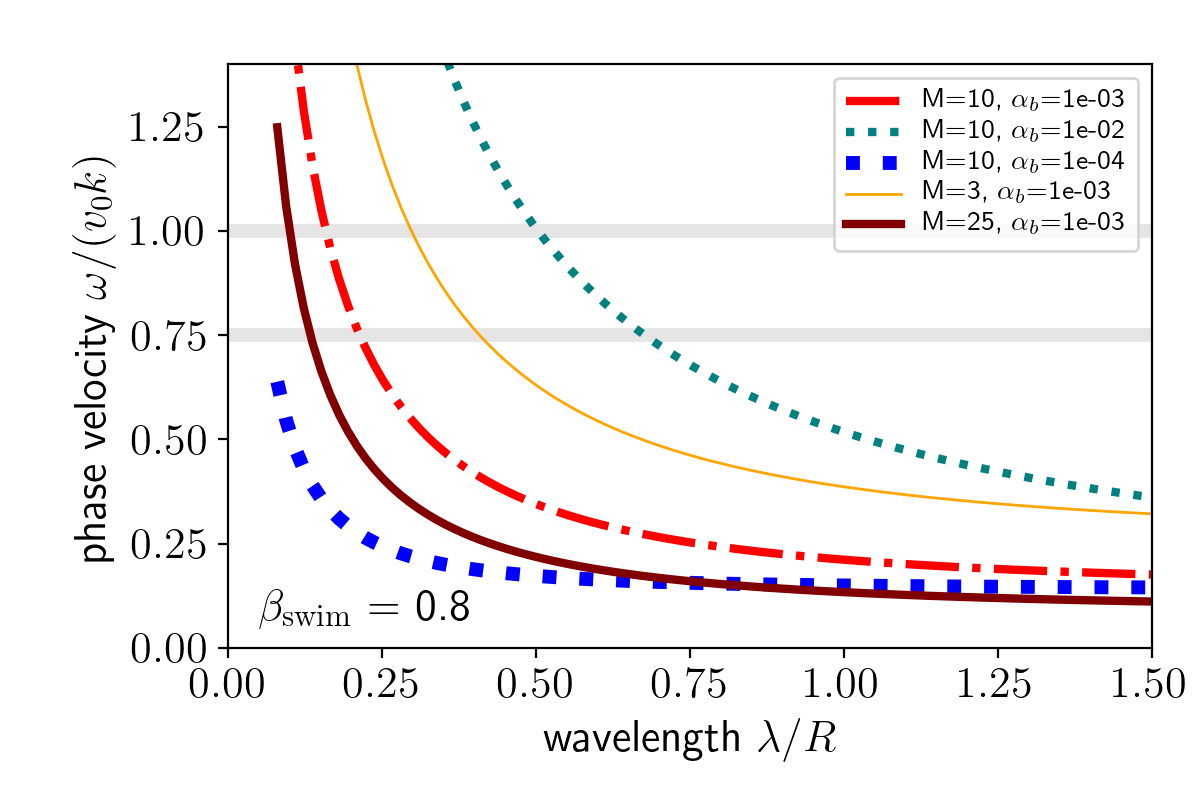} 
\caption{Phase velocities for waves on the boundary for a simple model that takes into account swim pressure
from boids.  These were computed using the dispersion relation in equation \ref{eqn:disp2}.
 a) No local boid swim pressure on the boundary. b) Local boid swim pressure set by $\beta_{\rm swim} = 0.8$.
 In both plots, orange solid, red dot-dashed, and maroon solid lines of increasing thickness have  mass ratio 
 $M_{\rm nodes}/M_{\rm boids} = 3,10,25$, respectively, and bending moment 
 $\alpha_{\rm bend}/(M_{\rm boids}  v_0^2 R) = 10^{-3}$.
Thin cyan and thick blue dotted lines have $\alpha_{\rm bend}/(M_{\rm boids}  v_0^2 R) = 10^{-2} $ and 
$10^{-4} $, 
respectively, and  mass ratio $M_{\rm nodes}/M_{\rm boids} = 10$.
Horizontal grey lines are at velocity  $v_0$ and $3/4 v_0$.  Wavelengths to the right
of where the curved lines cross a horizontal line have phase velocity below the value of the horizontal line.
If instability depends on matching boid speed to the velocity of waves on the boundary,
then smaller wavelengths are unstable for higher mass and more flexible boundaries. 
\label{fig:vphase}}
\end{figure}

In Figure \ref{fig:vphase}, we have plotted the phase velocity $\omega/(k v_0)$, computed using equation \ref{eqn:disp2},
as a function of wavelength for 
different boundary to boid mass ratios, bending moments and for two different values for the dimensionless
coefficient $\beta_{\rm swim}$.   The  values of boundary to total boid mass ratio and bending moments
are the same as used in our simulation series.  In Figure \ref{fig:vphase}a,   velocities are shown
for $\beta_{\rm swim} =0$.  This would be if the boids locally did not exert much pressure on the boundary that
is above or below a mean value.
In Figure \ref{fig:vphase}b,  velocities are shown  for $\beta_{\rm swim} = 0.8$.
Orange solid, red dot-dashed, and maroon solid  lines of increasing thickness have mass ratio 
 $M_{\rm nodes}/M_{\rm boids} = 3,10,25$, respectively, and bending moment 
 $\alpha_{\rm bend}/(M_{\rm boids}  v_0^2 R) = 10^{-3}$.
Thin cyan and thick blue dotted lines have $\alpha_{\rm bend}/(M_{\rm boids}  v_0^2 R) = 10^{-2} $ and 
$10^{-4} $,   respectively, and  mass ratio $M_{\rm nodes}/M_{\rm boids} = 10$.  We note that the phase velocities shown  in Figure \ref{fig:vphase}b  do not reach zero.  We suspect that instability is not caused by large $\beta_{\rm swim}$ which would give a negative right hand side to equation \ref{eqn:disp2} and so complex values for frequency $\omega$.  
In this sense, our estimates for the phase velocity do not support the model for boundary instability explored by \citet{Nikola16}.

Figure \ref{fig:vphase} illustrates  that higher boundary mass gives lower wave velocity on the boundary.  Likewise weaker  boundaries, (with lower
$\alpha_{\rm bend}$) have  lower  wave velocity.     The trends we see in Figure \ref{fig:mont}, showing
that corrugation wavelengths decrease with increasing boundary mass and decreasing
bending moment,  are  matched  by the trends we see in wave velocity.  This suggests that the instability
on the boundary grows when the wave speed on the boundary is similar to boid speed.
Horizontal grey lines on Figure \ref{fig:vphase}b show constant velocities.   Wavelengths to the right
of where the curved lines cross a horizontal grey line have phase velocity below the value of the horizontal line.
If instability depends on matching boid speed to the velocity of waves on the boundary,
then smaller wavelengths are unstable with higher mass and more flexible boundaries. 

Using our dispersion relation in equation \ref{eqn:disp2}, the wavevector  that gives $\omega = k v_0$
(and matching wave phase velocity to boid speed) is
\begin{align}
 k_{crit}  &=  \sqrt{   \frac{\mu v_0^2 - T(1-\beta_{\rm swim})}{\alpha_{\rm bend}}} 
 .
 \end{align}
 For 
 $M_{\rm nodes} > M_{\rm boids}$ and  the regime giving
 us interesting boundary morphology, the critical wave vector 
 \begin{align}
 k_{crit}  R  & \approx \sqrt{ \frac{R v_0^2 M_{\rm boids}}{2 \pi \alpha_{\rm bend}} }
    \sqrt{ \frac{M_{\rm nodes}}{M_{\rm boids}}  }. \label{eqn:kcrit}
 \end{align}
 In terms of a critical wavelength $\lambda_{crit} = 2 \pi/k_{crit}$,
 \begin{align}
\frac{ \lambda_{crit}}{R} 
 \approx 0.16 \left(\frac{ \alpha_{\rm bend} }{10^{-3} M_{\rm boids} R v_0^2}\right)^\frac{1}{2} 
\left( \frac{10}{ M_{\rm nodes}/ M_{\rm boids}}\right)^\frac{1}{2}.
\end{align}
The scaling and approximate values for the critical wavelength are consistent with the wavelengths
giving phase velocity of $v_0$ shown in  Figure \ref{fig:vphase}.
 
As long as the coefficient giving swim pressure strength $\beta_{\rm swim}  <1$,
the dispersion relation in equation \ref{eqn:disp2} always gives real frequencies $\omega$ when
the wavevectors are real.
Only wavelike solutions would be present and perturbations on the boundary
would not grow. If the dispersion relation has regions where 
 frequency $\omega$ is complex for real $k$, then perturbations at these wavelengths would
 grow exponentially giving instability on the boundary.
If the $k^2$ term in the dispersion is negative then there is an instability at small wavelengths. This is the setting discussed by \citet{Nikola16} for instability of a filament embedded in a medium containing self-propelled particles.
A modified form for the swim pressure might give a larger negative term in the dispersion relation and show instability.  

Using a linearized version of  Bernoulli's equation,
a two-dimensional incompressible fluid approximation for  boids moving  at $v_0$ would give 
 boid pressure perturbation  with amplitude
$p_k \propto M_{\rm boids}  (\omega - kv_0)^2/k$  for a perturbation $ \propto e^{i(\omega t - kx)}$ on the
boundary.
However  unstable regions in the dispersion relation then occur at  larger wavelengths
for  heavier  boundaries which is 
opposite to what is seen in our simulations (see Figure \ref{fig:mont}c).   A model where swim pressure is proportional
to boid density and boid density is proportional to the local boundary curvature (e.g., \cite{Fily2014}) 
also would predict  this trend that is {\it not} consistent with our simulations.   If the local swim pressure
is large and $\beta_{\rm swim} >1$ in equation \ref{eqn:disp2}, unstable
regions would also give this incorrect trend.  These types of instability models also predict rapid growth rates for the instability,
also in contradiction to what we see in the simulations, where  corrugations in the boundary 
take 5 to 10 crossing times $t_R$ to grow.   

The models discussed in the previous paragraph and equation \ref{eqn:pswim} (and by \cite{Nikola16}) have boid swim pressure perturbations,  exerted on the boundary, that are in phase with the boundary perturbation.  However, we see a difference in the boid motions between leeward and windward sides of corrugations in our simulations.    This is most
extreme for the massive boundaries on the right hand side of Figure \ref{fig:mont}c (third row) where boids are pushed outward toward the center of the enclosed region after they pass  a convex region of the boundary.  The difference between leeward and windward sides in the boid  motions implies there is an asymmetry in the response of the boids to perturbations in the boundary.
The response of the boids slightly lags behind the perturbation, giving a phase shift in the pressure response.

We consider a model where the boid swim pressure
is slightly out  phase with a small  perturbation on the boundary.  For a perturbation $ \propto e^{i(\omega t - kx)}$
on the boundary, we assume that the sign of the phase shift depends  on  $\bar v - \omega/k $ 
where  $\bar v$  is the mean speed of boids that are  next
to the boundary.  We approximate $\bar v \sim v_0$ even though the mean speed $\bar v$ is  usually  lower than $v_0$ because the boids are slowed by bouncing against the boundary.    
The  phase shift gives an additional complex component to the amplitude of the boid pressure perturbation
$p_{\rm swim,k}$.
We assume that the phase shift in boid pressure is in the same form as equation \ref{eqn:pswim}, contributing
a complex component
\begin{equation}
 {\rm Im} (p_{\rm swim,k}) =  i \delta_{\rm lag} Tk^2 {\rm sign} ( k v_0 - \omega)
 \end{equation}
 to the swim pressure perturbation amplitude.  
 Here $\delta_{\rm lag}$ is a small dimensionless parameter describing the size of the lag.
 Modifying equation \ref{eqn:disp2},
 the resulting dispersion relation is 
 \begin{equation}
 \omega^2 = \frac{\alpha_{\rm bend}}{\mu} k^4 +  \frac{T}{\mu} k^2 \left(1 - \beta_{\rm swim} 
 + i \delta_{\rm lag} {\rm sign} \left( k v_0 - \omega \right) \right)
 . \label{eqn:disp3}
\end{equation}
Assuming that the parameter $\delta_{\rm lag}$ is small, we find that the perturbation only grows if
the imaginary term on the right hand  is positive.   
An instability is present if  $v_0 > \omega/k$, so
 only boundaries with slow wave speeds  would be unstable
to the growth of corrugations.    As heavier boundaries have slower bending wave speeds, the delay would account for the relation between corrugation and boundary  mass we see in Figure  \ref{fig:mont}c.


With small $\delta_{\rm lag}$, we estimate an instability growth rate
from the imaginary component of the frequency
\begin{equation}
\gamma(k) = {\rm Im} (\omega)  \approx  \frac{\delta_{\rm lag} T k^2}{2\mu {\rm Re}(\omega(k))}.
\end{equation}
Unstable perturbations would have amplitudes that increase exponentially with time,  $\propto  e^{\gamma(k) t}$.
While all wavelengths larger than the critical one $\lambda_{crit}$, (where wave speed matches boid speed)
would be unstable (due to the sign of the phase lag), the growth rate
is maximum near the smallest unstable wavelength which is the critical one.  
Using equation \ref{eqn:kcrit} for the critical wavevector,
we estimate the  the growth rate for this wavelength, 
\begin{align}
 \gamma (k_{\rm crit}) t_R  &\approx  \frac{T \delta_{\rm lag}} {2 \mu v_0^2} k_{crit} R 
   \approx \frac{\delta_{\rm lag}}{2 }  
  \sqrt{ \frac{R v_0^2 M_{\rm boids}}{2 \pi \alpha_{\rm bend}} }   \sqrt{ \frac{M_{\rm boids}} { M_{\rm nodes}} } 
  \nonumber \\
  & \approx 2 \delta_{\rm lag} 
  \left(\frac{10^{-3}  M_{\rm boids} R v_0^2}{ \alpha_{\rm bend} } \right)^\frac{1}{2} 
\left( \frac{10}{ M_{\rm nodes}/ M_{\rm boids}}\right)^\frac{1}{2}.
   \label{eqn:grow}
\end{align}

\begin{figure}
\includegraphics[trim=5 5 0 5,clip,width=3.0in]{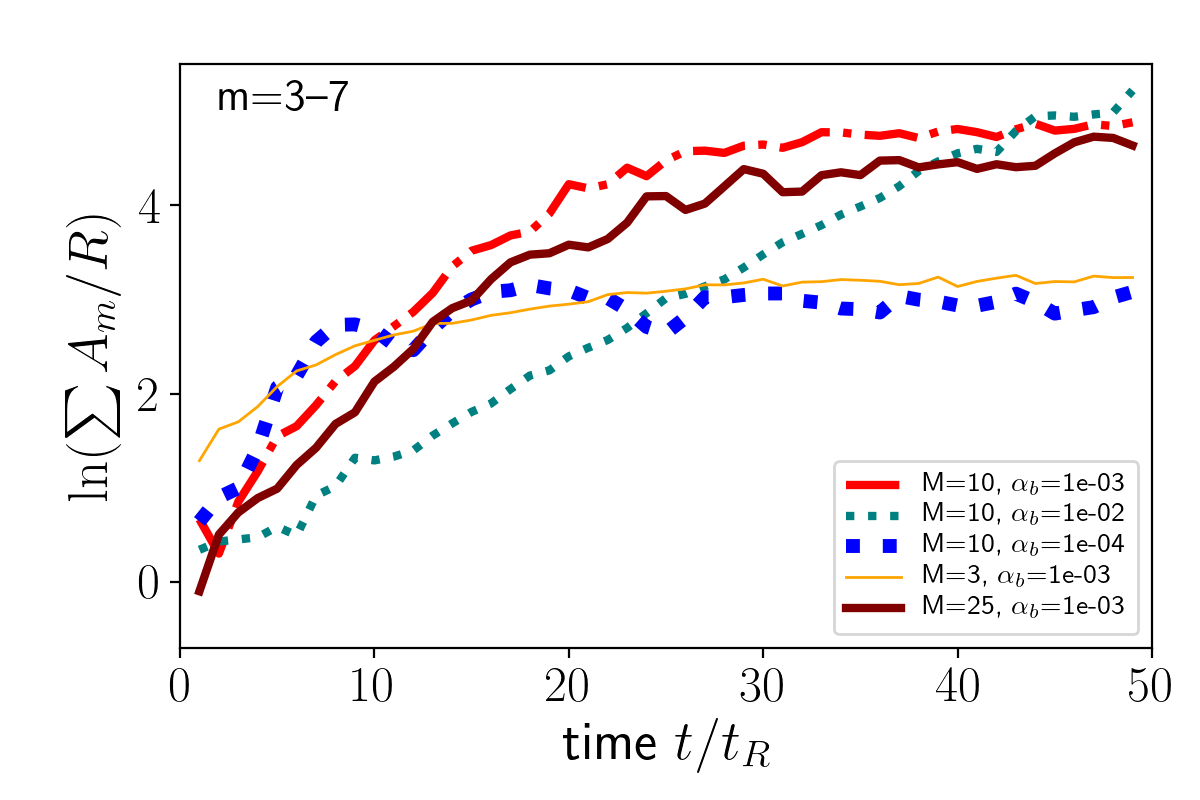} 
\includegraphics[trim=5 5 0 5,clip,width=3.0in]{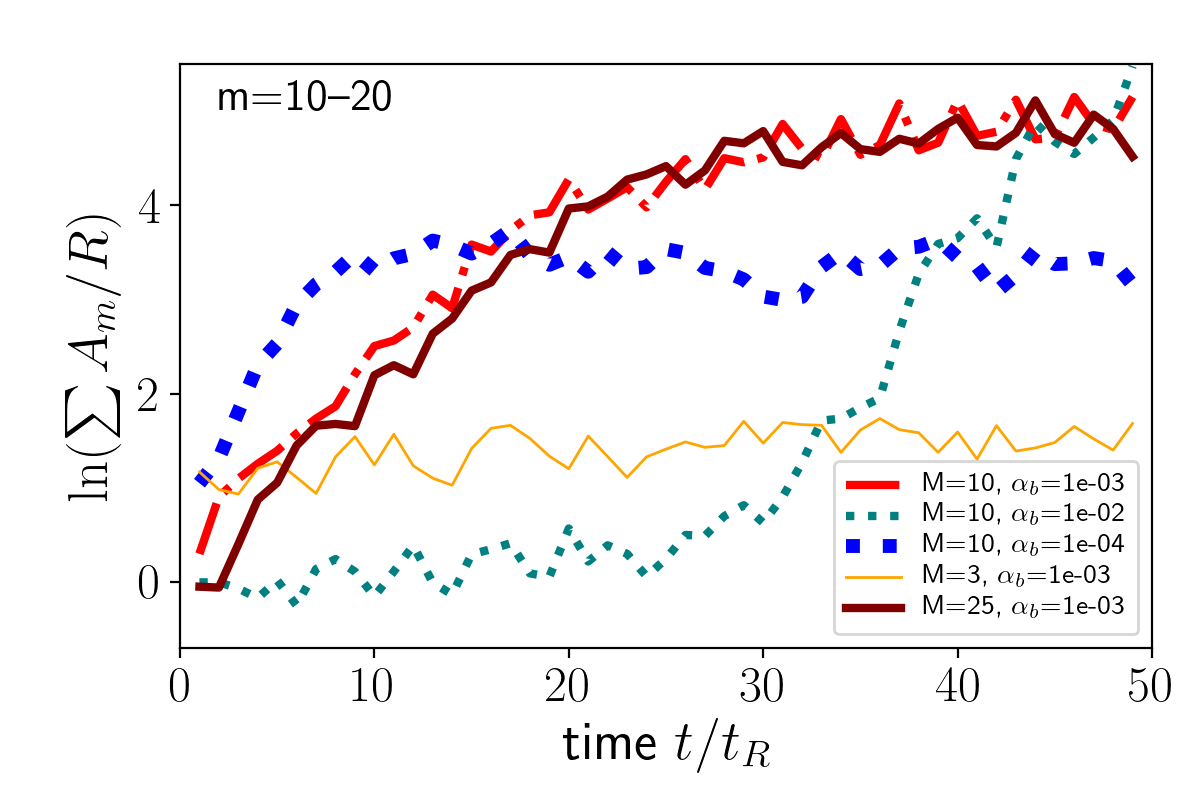} 
\caption{
The log of a sum of Fourier amplitudes measured from the boundary for 5 different  simulations.
a) Using the $m=3$ to 7 Fourier amplitudes.
b) Using the $m=10$ to 20 amplitudes.
The lines types and parameter choices are the same as in Figure \ref{fig:vphase}.
The simulations have parameters the same as the first and third series of simulations
listed in Table \ref{tab:loops} except they have specific boundary to boid mass ratios and bending moments that are shown in the legends. 
\label{fig:Am}}
\end{figure}

We can test this phase-lag instability model by examining the rate that boundary perturbations grow in our simulations.  
In 5 simulations we measure Fourier amplitudes $A_m(t)>0$ as a function of time, where integer $m$ gives the angular frequency   of radius $R(\theta,t) = \sum_m A_m(t) \cos (m\theta + \phi_m(t))$  as a function of angle 
$\theta$ along the boundary.   For example, a triangular perturbation gives an amplitude $A_3$.
The angle $\phi_3$ determines the orientation  of the triangular perturbation. 
The 5 simulations have parameters taken from Table \ref{tab:common}
and Table \ref{tab:loops} but with the boundary to boid mass ratio
and bending moments chosen to be the same as the phase velocities plotted in Figure \ref{fig:vphase}.
These simulations are the part of the first and third series listed in Table \ref{tab:loops}
and shown in the first and third rows of Figure \ref{fig:mont}.
In Figure \ref{fig:Am}a, we plot $\ln (\sum_{m=3}^7 A_m/R) $ as a function of time
and  in Figure \ref{fig:Am}b we plot $\ln (\sum_{m=10}^{20} A_m/R) $.  Lines have the same
colors and styles as in  Figure \ref{fig:vphase}.

Figure \ref{fig:Am} shows that corrugation growth rates are faster with lower
values of bending moment (comparing blue dotted, red dot-dashed and thin teal dotted lines), 
as expected from equation \ref{eqn:grow}.
The inverse dependence  of growth rate on boundary to total boid mass ratio
is less evident, but the mass ratio varies by a factor of about 3 rather than 10 as for
the bending moment.  The low mass boundary only grows larger wavelength perturbations (with  lower
Fourrier index $m$) and the growth rate is slower than for the higher mass boundaries
with the same bending moment (comparing thin orange to thick red and maroon lines).
The trends we see in Figure \ref{fig:Am} are consistent with those predicted by equation \ref{eqn:grow}.

We use our numerically measured growth rates to estimate the size of the pressure lag.
In equation \ref{eqn:grow} we have  estimated the  growth rate of the critical wavelength
for the  mass  ratio 10 and $\alpha_{\rm  bend}/(M_{\rm boids}  v_0^2 R)   = 10^{-3}$ simulation which is shown
with a dot-dashed red line in  Figure \ref{fig:Am}.  The slope of the red  line gives a  growth rate  of $\gamma t_R \sim 0.2$.
Equating this to the growth rate  in  equation \ref{eqn:grow} we estimate 
$\delta_{\rm lag} \sim 0.1$. 
The required  lag for the pressure is small enough to be consistent
with the appearance of the simulations.  This implies that a small delay in boid response  moving
over boundary perturbations can account for the instability on the boundary.


Throughout the discussion in this section we have assumed that tension on the boundary was
that estimated by equation \ref{eqn:tension}.   However if the boid separation is shorter than  the 
repel distance, $\sqrt{\frac{\pi R^2}{N_{\rm boids}}} < d_{\rm repel}$,   then there is additional  tension on the boundary because the boids are pushed against the boundary by their repulsion.   An increase in tension increases the wave speed and would 
reduce the wavelength of corrugations on the boundary.   The second series of simulations
shown in Figure \ref{fig:mont}b (second row) is in this regime and shows weaker boundary perturbations.
Comparison of this simulation to that with identical parameters but lower repel distance $d_{\rm repel}$  (Figure \ref{fig:mont}a, top row) shows that the corrugations in the higher tension simulations tend to be shorter wavelength, confirming our expectation. 
A single Fourier perturbation tends to dominate in these simulations, but we lack an explanation for this phenomenon.

What accounts for the size of the phase lag parameter $\delta_{\rm lag}$?  The phase lag may be due
to the time it takes other boids to push near-boundary boids back onto the boundary.   This time might be governed
by the strength of the interboid repel force.
We have noticed that a weaker repel force $U_{\rm repel}$ gives larger density contrasts in the boids.
We would expect this to give a larger asymmetry between windward and leeward sides of corrugations in the boid distribution, leading to faster corrugation growth rates and larger amplitude corrugations but not necessarily a change in the wavelengths that are unstable.  
However, in Figure \ref{fig:mont}e (fifth row), 
the simulations with lower $U_{\rm repel}$ do seem to have smaller wavelength corrugations
and with larger $U_{\rm repel}$,  the boundary instability
is suppressed.   The variation in the wavelengths of instability must be due to another cause, 
perhaps because changing $U_{\rm repel}$ also affects boid density near the boundary and the pressure related tension on the boundary, which in turn affects the speed of boundary waves.
Boids are slowed down near the boundary and if the mean speed depends on $U_{\rm repel}$, this too could affect the wavelengths that are unstable.
We lack a straightforward way to predict the delay parameter, $\delta_{\rm lag}$.
Better understanding of the boid's continuum dynamics near the boundary  may make it possible to predict the phase lag from the repel force law and mean boid number density.

In summary, we have explored simple models for boid swim pressure, exerted onto the boundary, that would give instability  on the boundary.    A model with boid swim pressure dependent
on the boundary curvature and slightly lagging its corrugations is most successful  at matching  
sensitivity of boundary corrugation wavelength to boundary mass and bending moment and the corrugation
growth rates. 
Perturbations on the boundary that move with wave speed slower than but near the boid speed are most likely to grow and this determines the wavelengths that grow on the boundary.  

\section{Summary and Discussion}
\label{sec:sum}

We have carried out a numerical exploration in 2-dimensions of self-propelled particles with alignment and repelling forces that are enclosed in a flexible elastic loop.  We primarily find three types of long lived states: a stochastic gas-like state, a solid-like or jammed bullet state where the boids align and push the boundary in a single direction and 
rotating or circulating states.
The gaseous and circulating states resemble those exhibited by unconfined unipolar self-propelled particles with cohesive or attractive interactions \citep{Levine2000,Touma2010}. The solid-like state resembles the jammed state seen in simulations of confined soft repulsive self-propelled particles at high density \citep{Henkes2011} and the moving droplets seen in simulations of unconfined self-propelled particle with strong cohesion  \citep{Gregoire2004,Touma2010}.  
We recover these three types of states without cohesion due to the confining nature of the boundary.  

The most of interesting and novel of the states exhibited by our simulations are the circulating states as they include rotating ovals and sprocket shaped and irregular or ruffled  boundaries.   Oval shaped boundaries are similar to those
seen in simulations of  non-aligning stochastically perturbed self-propelled particles \citep{Paoluzzi2016,Wang19}.   
The ruffled or sprocket shaped rotated boundaries  
mimic the rotating ratchet that was achieved by placing a rigid ratchet in an solution of active particles \citep{Sokolov2010,DiLeonardo2010,Angelani2011}, but here the collective motion of the
self-propelled particles and instability on the boundary drive the rotation.
The instability is likely mediated by boid pressure inhomogeneities, as predicted by \citet{Nikola16}.
However, the instability is most noticeable in the simulations with more massive and flexible boundaries.
The wavelength of corrugations on the boundary is near the wavelength of elastic waves on the boundary 
that  have phase velocity equal to the particle swim speed.   We suspect that the instability depends on a lag between boid  swim pressure exerted on the boundary and boundary shape perturbations.  In this sense our instability model 
differs from that by \citet{Nikola16} who lacked a phase lag in their instability model.

It may be possible to devise an experiment giving an instability on a flexible boundary that is mediated by active particles. 
Here we considered a uniform loop boundary, but a boundary could be designed to be more flexible in one region than another.  
For example, if the instability is fast, waves might be excited on one side of a loop, making it possible to fix the other side to another surface. 
States with rotating or fluttering boundaries might be used to generate fluid flow or vorticity or to create a swimmer.  These artificial mechanisms  could more efficiently use power from self-propelled particles as the particles are in proximity to the moving boundary rather than distributed in a solution, though  providing the particles with an energy source for propulsion could be more difficult as their fuel must be stored within or cross the boundary.

In this study we ignored stochastic perturbations and cohesion in  the self-propelled particles
and the hydrodynamics of the medium in which the self-propelled
particles move.    
Phase diagrams for classes of DADAM tend to scale with the ratio of density to noise strength,
with noisier systems more likely to display disordered phases \citep{Chate2019}.
Our simulations were restricted to a few hundred boids. 
Future studies could extend and vary the physical model and explore dynamics in three dimensions.
Future work could also explore other types of active materials that are enclosed by flexible boundaries, such as active self-propelled rods (e.g., \cite{Kaiser2012,Bar2019}) or active nematics (e.g., \cite{Ramaswamy2003,Sanchez2012,Marchetti2013,Thampi2014,Chate2019}).
With unipolar self-propelled particles,  we did not see long lived bending oscillations. Perhaps other types of active materials enclosed in a flexible boundary could exhibit this type of phenomena.

\FloatBarrier
 
\vskip 0.3 truein

\begin{acknowledgments}

We thank Steve Teitel and Randal C. Nelson for helpful discussions.
This material is based upon work supported in part  by NASA grant 80NSSC17K0771, National Science Foundation Grant No. PHY-1757062, and  National Science Foundation Grant No. DMR-1809318.

\end{acknowledgments}

\appendix

\section{Numerical Implementation}
\label{sec:num_imp}

All boundary node masses are equivalent and all boid masses are equivalent, however node mass
is usually not equal to boid mass.  The total number of boids and nodes remains fixed during the simulation.
For visualization,  we translate the viewing window so that it is centered on the center of mass of the boundary. 

\subsection{Initial conditions}
\label{sec:init}

The  simulations are initialized with boids initially confined within a circle with radius of 0.9 the initial boundary radius, $R$.
Boids are initially uniformly and randomly distributed within this 
this circle.  We explored two types of initial conditions for the boids, an initially rotating
flock and a nearly stationary flock.   
In both cases we also added a small initial random velocity, uniformly distributed in angle, of size 0.1 $v_0$, 
where $v_0$ is the boid swim speed.
The rotating swarm has boids initially  rotating about the boundary center at a velocity of 0.8 $v_0$.  
Circulating initial conditions are chosen when we study the circulating states, whereas
random initial conditions without mean rotation are chosen when we  study the transitions between gaseous-like, circulating
and  jammed states.


The boundary nodes are initially placed in a circle of radius $R$, 
equally spaced and at zero velocity.   
Springs between neighboring nodes are initially set to their rest length and all springs have
the same spring constant.  The bending moment does not vary as a function of position
on the boundary.

\subsection{Units}
\label{sec:units}

We work in units of boid speed $v_0$,  initial boundary radius, $R$ and total boid mass 
$M_{\rm boids}$.  A unit of time is  
\begin{equation} t_R \equiv R/v_0 ,\label{eqn:t_R}
\end{equation}
which is the time for a lone boid moving at $v_0$ to cross the  radius  $R$ of the boundary. 
After choosing these units, the free parameters are the total boundary mass  $M_{\rm nodes}$ which is also
the boid to boundary mass ratio, the number of nodes and boids $N_{\rm nodes}$ and $N_{\rm boids}$,  the alignment force strength and length scale, $\alpha_{\rm align}$ and $d_{\rm align}$, the repel force strength and length scale $U_{\rm repel}$ and $d_{\rm repel}$, 
the bending moment, 
$\alpha_{\rm bend}$, the node damping parameter $\gamma_{\rm damp}$, the node-boid interaction strength and length scale, $F_{\rm interact}$ and $d_{\rm interact}$, and  the spring constant $k_s$. 
To run a simulation we also require a time step $dt$, which is fixed during the simulation, and a maximum length of time $t_{\rm max}$ to integrate.
This is a large parameter space, but not all combinations of these parameters necessarily affect the collective dynamics or are in regimes that are physically interesting or could be realized numerically.
As long as number of nodes is high enough that the boids are confined and they smoothly interact
with the boundary, the dynamics should not depend on the number of nodes in the boundary or the parameters
describing the boid/node interactions.  The springs are used to set the boundary length so the spring
constant should not affect the dynamics. The dynamics could depend upon the number  and mass of boids as the swim pressure, or pressure exerted by boids on the boundary, depends on their number density.  

\subsection{The time step}
\label{sec:dt}

The speed of compression waves traveling in a linear mass/spring chain  is
\begin{equation}
    v_c = \sqrt{\frac{k_s}{m_{\rm node}}} \Delta s = \sqrt{\frac{k_s}{m_{\rm node}}} \frac{2 \pi R}{N_{\rm nodes}}.
\end{equation}
For numerical stability, a CFL-like condition for the time step is that it must be less than the time it takes a compression wave to travel between nodes or 
\begin{equation}
dt < \sqrt{\frac{m_{\rm node}}{k_s}}. 
\label{eqn:dt1}
\end{equation}
In the continuum limit,
equation \ref{eqn:BB} gives a dispersion relation for bending waves equivalent to that from Euler-Bernoulli beam theory
\begin{equation}
  \omega^2 = \frac{ \alpha_{\rm bend}}{\mu} k^4, 
  \label{eqn:waverelation}
\end{equation}
where $\alpha_{\rm bend}$ is the bending moment or flexural rigidity,  $\mu =m_{\rm node}/\Delta s$ is the linear mass density, $\omega$ is angular wave frequency and $k$ the wavevector. 
The simulation time step should be chosen so that small corrugations in the boundary are not numerically unstable.  Taking the wave speed for wavevector $k = 1/\Delta s$, from the node separation, a condition on the time step for numerical stability is
\begin{equation}
    dt < \sqrt{ \frac{m_{\rm node}}{\alpha_{\rm bend} \Delta s} } (\Delta s)^2. \label{eqn:dt2}
\end{equation}
The time step should be shorter than the time it takes a boid to travel between boundary nodes,
the mean distance between boids, and
the repel, align and boundary interaction distances,
\begin{equation}
dt < {\rm min} \left( \frac{ \Delta s}{v_0}, \frac{1}{v_0}   \sqrt{ \frac{\pi R^2}{N_{\rm boids}}}, \frac{d_{\rm repel}}{v_0}, \frac{d_{\rm align}}{v_0}, \frac{d_{\rm interact}}{v_0} \right)
 . \label{eqn:dt3}
\end{equation}
We chose time step to satisfy   equations \ref{eqn:dt1}, \ref{eqn:dt2}, and \ref{eqn:dt3},  with equation \ref{eqn:dt2}  usually the most restrictive.

The springs are present to keep the boundary length nearly constant.  We would like the springs to be strong enough that the choice of spring constant does not affect the  simulation collective behavior.  
Because they must turn,
boids circulating near a circular boundary exert a pressure on the boundary.
The force per unit length on the boundary 
is $p \sim M_{\rm boids}\frac{v_0^2}{R} \frac{1}{2 \pi R}$.  This  pressure is balanced by a 
tension in the boundary (sometimes called wall tension and related to hoop stress) 
that depends on the curvature of the boundary,   $p \sim {T/R} $.  Balancing these two
estimates, we estimate the tension on the boundary 
\begin{equation}
T \sim M_{\rm boids} \frac{v_0^2}{R} \frac{1}{2 \pi }.   \label{eqn:tension} 
\end{equation}  
This tension can stretch each spring by $\delta x$ from its rest length, giving tension $T = k_s \delta x$.
The spring strain is $\epsilon = \delta x/\Delta s $ with spring rest length $\Delta s = 2 \pi R/N_{\rm nodes}$.  Setting tension from wall strain equal to that from spring tension, we solve for the spring strain to give a dimensionless parameter
\begin{equation}
 \epsilon_{ks} \equiv \frac{M_{\rm boids} v_0^2}{(2 \pi R)^2 } \frac{N_{\rm nodes}}{k_s}. \label{eqn:epsilon_ks}
 \end{equation}
As long as this parameter is small, the springs should remain near their rest length and the choice of spring constant should not affect the behavior of the simulations.
We ensure that our spring constant $k_s$ is large enough that $\epsilon_{ks} < 1$ is satisfied.

\subsection{Other constraints on parameters}
\label{sec:other}

The boundary/boid interaction should primarily cause boids to reflect off the boundary.  The acceleration on the boids from the boundary nodes should exceed the interboid repel force
\begin{align}
 \frac{ F_{\rm interact}}{m_{\rm boid}}\frac{d_{\rm interact}}{\Delta s}  \gtrsim \frac{U_{\rm repel}}{d_{\rm repel}}
\end{align}
where the factor $d_{\rm interact}/\Delta s$ describes the number of nodes that push away a single boid as it approaches the boundary. 
We also require internode distance to be similar or less than the boundary interaction
distance, $\Delta s \lesssim d_{\rm interact} \ll R$.   The interaction force  should not be
so large that  boids on the  boundary move a  large distance  during  a single  time step,  giving an upper bound  
\begin{align}
\frac{F_{\rm  interact}}{m_{\rm  boid}} \frac{dt}{v_0} \lesssim 1.
\end{align}
We maintain these conditions so that the parameters describing the boid/node interaction force should not significantly affect the boid collective behavior.
We have halved the time step and we doubled the spring constant to check that these did not affect our simulations. 
We repeated simulations to check
that boid distribution and boundary morphologies look similar at the end.  There
is  sensitivity to initial conditions with some simulations freezing or jamming in a bullet-like state
and others with the same parameters remaining in a circulating state.  This is discussed in
more detail in section \ref{sec:phenomena}.

If the interboid alignment force is too weak, then many boid crossing travel times would be required for collective phenomena to develop.  We maintain alignment strength $\alpha_{\rm align} t_R > 1$ 
so that self-propelled particles align on a timescale shorter than the travel time across the enclosed region. 
This condition also ensures that transient behavior decays within a few dozen domain travel times, $t_R$.   
Likewise we keep the repel strength divided by the square of the swim speed $U_{\rm repel}/v_0^2$ to be of order 1 so that the boids effectively repel one another during a simulation extending a few dozen crossing times $t_R$.
There is some degeneracy between alignment strength $\alpha_{\rm align}$ and distance $d_{\rm align}$ in how these parameters affect collective behavior as both
affect boid alignment.   There is also a degeneracy between repel strength $U_{\rm repel}$ and distance $d_{\rm repel}$  as both parameters determine interboid repulsion.  Consequently we usually fix the alignment and repel strengths 
$\alpha_{\rm align}$  and $U_{\rm repel}$, and vary their length scales $d_{\rm align}$ and $d_{\rm repel}$  in our numerical exploration of collective phenomena.

The damping parameter $\gamma_{\rm damp}$ mimics friction or viscous interaction with a background substrate or fluid. 
If the damping parameter $\gamma_{\rm damp} t_R \gg 1$ then the boundary is over-damped and will not be sensitive to boid pressure.   If $\gamma_{\rm damp} t_R $ is extremely small, then circulating boids within the boundary will cause the boundary to rotate, eventually matching the boid rotation speed.   
We set $\gamma_{\rm damp} t_R  = 0.1$, an intermediate value, so that transient behavior will decay within a few dozen crossing times.

To allow transient behavior to decay, we run each simulation for  $t_{\rm max} = 50 t_R$. 
We show in section \ref{sec:inst} that the growth of structure on the boundary usually saturates by this time. 

\subsection{Code repository}
\label{sec:repo}

We checked our classification of collective behavior and phenomena with two independently written codes.
One version is written in C, uses an openGL display and nearest neighbor searches are accelerated with a 2D quad-tree search algorithm based on the Barnes-Hut algorithm \citep{Barnes1986}. 
This code can be found here: \url{https://github.com/jsmucker/boids-in-a-boundary}.
Another version of our code is written in Javascript using the \texttt{p5.js} library
(see \url{https://p5js.org/}).  
This code displays in a web-browser and nearest neighbor searches are not accelerated.  
This code is available on github at \url{https://github.com/aquillen/boids_in_a_loop}.
The figures in this manuscript were made with the Javascript code.

\newpage
\bibliography{boids}


\end{document}